# Enhancement of spin-wave nonreciprocity and group velocity in a low-wavenumber regime

Shion Yoshimura[1], Shugo Yoshii[1,2], Ryo Ohshima[1,3] and Masashi Shiraishi[1,3*]

[1]Department of Electronic Science and Engineering, Kyoto University, Nishikyo-ku, Kyoto 615-8510, Japan

[2]Blackett Laboratory, Imperial College London, London, SW7 2AZ, United Kingdom

[3]Center for Spintronics Research Network, Institute for Chemical Research, Kyoto University, Uji, Kyoto 611-0011, Japan

[*] Masashi Shiraishi, Email address: shiraishi.masashi.4w@kyoto-u.ac.jp

**Abstract**

Nonreciprocity of spin waves is essential for components such as magnetic isolators and circulators used in spin-wave-based computing. A ferromagnetic (FM) bilayer exhibits significant frequency nonreciprocity and has attracted attention in recent years. Prior research on bilayers has predominantly focused on the high-wavenumber regime, where spin waves display significant nonreciprocity and are accessible through Brillouin light scattering (BLS). However, the dynamics at low wavenumbers ($k < 5$ rad/μm), which enable rapid magnon propagation, have yet to be thoroughly investigated. We investigate spin-wave propagation in the bilayer using coplanar waveguides (CPWs) and demonstrate that increasing the bilayer thickness enhances nonreciprocity even at low wavenumbers, which leads to the high group velocity originating from the Damon-Eshbach (DE) mode. These findings establish design principles for high-speed, low-loss spin-wave-based information processing.

# I. INTRODUCTION

Spin waves are the elementary excitations of magnetic systems. Their low power consumption and high excitation frequencies (GHz to THz) make them promising new information carriers [1–3]. The development of spin-wave devices, such as isolators and circulators, is a crucial step for efficient spin-wave-based computing, because they prevent the backflow of spin waves and propagate them in a particular direction [4,5]. One approach for developing these functionalities is to leverage the frequency nonreciprocity of spin waves (counterpropagating spin waves have different frequencies), which is attributed to symmetry breaking in the vertical direction of ferromagnetic (FM) thin films [6]. Various mechanisms contributing to nonreciprocity have been reported, including interfacial Dzyaloshinskii-Moriya interaction [7–9], magnetic surface anisotropy [10,11], and dipolar interaction [6,12–17].

An FM bilayer, composed of two ferromagnets with different saturation magnetizations, exhibits large nonreciprocity of spin waves caused by the strengths of dipolar coupling between the Damon-Eshbach (DE) mode and the perpendicular standing spin-wave (PSSW) modes, depending on the propagation direction. Consequently, it has been extensively studied in recent years, and the dispersion relations of the spin waves in the FM bilayer have been theoretically and experimentally explored [13–15]. However, previous studies focused on nonreciprocity in a high-wavenumber regime ($k > 5$ rad/µm) caused by the higher frequency nonreciprocity and the limitations of Brillouin light scattering (BLS), which only accesses spin waves with a large amplitude at the top surface of the FM bilayer [14,15]. Spin waves in a low-wavenumber regime ($k < 5$ rad/µm) typically have higher group velocities than those in a high-wavenumber regime, which is pivotal for practical spin-wave devices. In contrast, the magnitude of the frequency nonreciprocity in a lower-wavenumber regime remains elusive. Therefore, the approach to enhancing the nonreciprocity by exciting the spin waves with high wavenumbers results in a bottleneck for high-speed information processing and leads to increased transmission loss [18].

We observe spin waves with considerably large frequency nonreciprocity and high propagation velocity in this study. These characteristics are achieved by increasing the bilayer thickness, which induces hybridization between the DE mode and PSSW modes at lower wavenumbers, where the DE mode exhibits a high group velocity.

## II. THEORY

A theoretical model that serves as a fitting function and a theoretical analysis of spin waves is presented (Refs. [11,14] for the calculation method; calculation details are provided in the Appendix). We consider an FM bilayer depicted in Fig. 1(a), where the saturation magnetization, exchange stiffness parameter, and surface magnetic anisotropy constant of layer 1 (layer 2) are represented as $M_{s1}$ ($M_{s2}$), $A_1$ ($A_2$) and $K_s^{\text{bot}}$ ($K_s^{\text{top}}$), and the thickness of each layer is equal and denoted as $l/2$. The Landau-Lifshitz (LL) equation is used to derive the dispersion relations and amplitudes of spin waves. The components of the effective magnetic field, dipolar field, exchange field, and anisotropy field are considered. The ratio of the exchange stiffness parameter to saturation magnetization is assumed to be equal in each layer, and is denoted as $\chi$ ($= A_1/M_{s1} = A_2/M_{s2} = \langle A \rangle / \langle M_s \rangle$) as in the previous work [14]. The equation is expanded using an orthonormal basis $\{S_0 \hat{\boldsymbol{x}}, S_0 \hat{\boldsymbol{z}}, S_1 \hat{\boldsymbol{x}}, S_1 \hat{\boldsymbol{z}}, \ldots, S_n \hat{\boldsymbol{x}}, S_n \hat{\boldsymbol{z}}\}$, where orthonormal basis functions $S_n(x)$ are denoted as

$$S_n(x) = \begin{cases} \sqrt{1/l} & \text{if } n \text{ is } 0, \\ \sqrt{2/l} \sin(n\pi x/l) & \text{if } n \text{ is odd}, \\ \sqrt{2/l} \cos(n\pi x/l) & \text{if } n \text{ is even}, \end{cases} \quad (1)$$

and each $S_n(x)$ is illustrated in Fig. 1(b). Consequently, the LL equation is represented as $i\Omega \boldsymbol{\eta} = C \boldsymbol{\eta}$, where $\Omega$ is the dimensionless angular velocity, $\boldsymbol{\eta}$ is the normalized amplitude vector with $2(n + 1)$ components and expressed as $\boldsymbol{\eta} = (\eta_x^0, \eta_z^0, \eta_x^1, \eta_z^1, \ldots \eta_x^n, \eta_z^n)^{\text{T}}$, and $C$ is the square matrix of $2(n + 1)$ rows and columns. Therefore, by calculating the eigenvalues and eigenvectors of $C$ numerically, the dispersion relations and normalized amplitudes of the spin waves in the FM bilayer are derived.

The dispersion relations of spin waves with the total thickness of 50 nm and 100 nm are depicted in Figs. 1(c) and 1(d) (the other parameters for the calculation are set to be $\mu_0 M_{s1} = 1$ T, $\mu_0 M_{s2} = 1.7$ T, $\chi = 12$ aJ/A, $K_s^{bot} = K_s^{top} = 0$ J/m$^2$ and $\mu_0 H_0 = 30$ mT), which enables the extraction of several physical features. First, the dispersion relations are governed by the dipolar coupling between the DE mode and the PSSW modes, and the asymmetry in the coupling strength endows the nonreciprocity (the latter modes are labeled as Mode I, Mode II, ... in order of increasing frequency to distinguish the spin-wave modes before and after coupling).

Second, the frequencies of the PSSW modes decrease as the total thickness increases. Because the exchange field is proportional to $\nabla^2 \boldsymbol{m}$ [19], excitation frequencies of $n = 1, 2, 3, ...$ component (Fig. 1(b)) reduce as increasing the thickness, and the frequencies of the PSSW modes, which are dominated by $n = 1, 2, 3, ...$ component, also decrease. The reduction in frequencies leads to coupling between the DE mode and the PSSW modes at lower wavenumbers, where the DE mode possesses a high group velocity.

Finally, the gradient of the DE mode increases with increasing film thickness, as suggested in Ref. [20]. Therefore, increasing the bilayer thickness exhibits a synergistic effect on achieving nonreciprocity with high propagation velocity.

## III. EXPERIMENT

Figure 2(a) illustrates a schematic of a fabricated device. The lithography-defined channel structures, consisting of a stack of SiO$_2$ (50 nm)/MgO (3 nm)/Co ($l/2$ nm)/Ni$_{80}$Fe$_{20}$ ($l/2$ nm), where $l = $ 50 nm or 100 nm, are deposited on a SiO$_2$/Si substrate using electron beam deposition (MgO, Co, Ni$_{80}$Fe$_{20}$) and radio frequency magnetron sputtering (SiO$_2$), where the MgO layer is equipped to prevent oxidation of the FM layers before sputtering the SiO$_2$. Coplanar waveguides (CPWs) of Au (100 nm)/Ti (3 nm) are fabricated on the SiO$_2$ insulating layer using electron beam lithography and electron beam deposition. The CPWs have a meandered shape as illustrated in Fig. 2(b) to enhance the selectivity of the wavenumber and excitation efficiency [21]. Six different types of CPWs were fabricated to investigate the dependence of nonreciprocity on the wavenumbers (the designed device parameters are listed in Table I). Figure 2(c) is a scanning electron microscope (SEM) image of CPW6.

A sketch of the measurement setup is depicted in Fig. 2(d). The CPWs on the FM bilayer are connected to a vector network analyzer (VNA, Keysight E5071C) via coaxial cables. The cables are terminated with probe needles electrically connected to the CPWs. Using the VNA, transmission spectra of spin waves are obtained from S-parameters ($S_{21}$ and $S_{12}$). The frequency range is set to be 300 kHz to 18 GHz with the power of 5 dBm. An external magnetic field is applied to the device in the $y$ direction and is swept from 0 mT to 100 mT.

## IV. RESULTS AND DISCUSSION

Figure 2(e) illustrates the measurement results of the Co (25 nm)/Ni$_{80}$Fe$_{20}$ (25 nm) sample equipped with CPW4, where the right (left) half of the color map corresponds to $S_{21}$ ($S_{12}$), the positive (negative) wavenumber. Because the raw data contain noise caused by electromagnetic waves, the measured S-parameters, $S_{ij}$, were baseline-corrected by subtracting the 0 mT ($|\Delta S_{ij}| = |S_{ij} - S_{ij}(0 \text{ mT})|$) to obtain $|\Delta S_{ij}|$. Furthermore, to remove noise from other sources (e.g., thermal noise), a low-pass filter was applied to the experimental data, resulting in the observation of the salient spin-wave signal. In Fig. 2(f), $|\Delta S_{21}|$ and $|\Delta S_{12}|$ are plotted as functions of frequency when the external magnetic field is 30 mT. The multiplicity of the maximum points can be attributed to the wavenumbers defined by the design of the CPWs [21] and the dispersion relations of several spin-wave modes. Therefore, to identify the wavenumbers and modes of an individual signal, the aforementioned theoretical model was used for the fitting, the procedure of which consists of three steps: (1) the spin-wave mode corresponding to a series of peaks extracted from the color map is identified by adjusting the wavenumber in the $z$ direction $k_z$, where the mode number was selected manually. The value of $k_z$ is estimated using the Fourier transform of the current density generated by the CPW [21]. (2) The peak positions in the spectra are fitted using six parameters: $k_z, M_{s1}, M_{s2}, \chi, K_s^{bot}$ and $K_s^{top}$, allowing determination of the values of $K_s^{bot}$ and $K_s^{top}$. (3) The fitting is repeated using only four parameters ($k_z, M_{s1}, M_{s2}$ and $\chi$), where the $K_s^{bot}$ and $K_s^{top}$ are fixed to the values obtained in the second step. This step is employed to reduce the uncertainties of the fitted parameters, which tend to increase significantly when all six parameters are fitted simultaneously.

The measured dispersion relations in Co (25 nm)/Ni$_{80}$Fe$_{20}$ (25 nm) (bilayer A) and Co (50 nm)/Ni$_{80}$Fe$_{20}$ (50 nm) (bilayer B) are depicted in Figs. 3(a) and 3(b), consistent with the theoretical calculation (the parameters are set to be $\mu_0 H_0 = 30$ mT, $\mu_0 M_{s1} = 1.01$ T, $\mu_0 M_{s2} = 1.77$ T, $\chi = 13.0$ aJ/A, $K_s^{bot} = 0.56$ mJ/m² and $K_s^{top} = -0.84$ mJ/m² in bilayer A, and $\mu_0 H_0 = 30$ mT, $\mu_0 M_{s1} = 1.02$ T, $\mu_0 M_{s2} = 1.63$ T, $\chi = 11.4$ aJ/A, $K_s^{bot} = 0.37$ mJ/m² and $K_s^{top} = -0.41$ mJ/m² in bilayer B. Detailed information on the fitted parameters is provided in Supplemental Material A [22]). These results imply that the CPW efficiently excites spin waves corresponding solely to the DE mode, which is the fundamental difference from BLS spectroscopy that excites spin waves localized near the top surface of a FM film [14,15].

The efficient excitation of the DE mode by the CPW is corroborated, as depicted in Figs. 3(c) and 3(d), where the amplitude of the uniform ($n = 0$) component of the magnetization along the $z$ direction, $|\eta_z^0|$, is visualized by the intensity of the red color. These figures reveal that the DE mode contains a high proportion of the uniform component. Because an alternating magnetic field generated by the CPW can be regarded as uniform in the vertical direction of FM thin films, the field strongly couples to the DE mode. Consequently, the CPW enables effective excitation and detection of the DE mode.

The central results of this work, i.e., the frequency differences $\Delta f (= f(+k_z) - f(-k_z))$ for bilayer A and bilayer B are depicted in Fig. 4. Although the maximum $\Delta f$ observed in bilayer A becomes greater than that in bilayer B in the high-wavenumber region, noticeable features are observed for bilayer B in the lower-wavenumber region. The group velocity $v_g$ of the spin wave in bilayer A is 2.56 km/s at $k_z$ = 7.48 rad/μm, whereas that in bilayer B is 6.28 km/s at $k_z$ = 3.24 rad/μm ($\Delta f$ = -1.05 GHz in both wavenumbers). Hence, the increase in the bilayer thickness enhances nonreciprocity significantly in the DE mode at low wavenumbers, resulting in large nonreciprocity with a high group velocity.

Based on the theoretical model, the thickness of the bilayer can be optimized to satisfy both large nonreciprocity and high group velocity in magnon transport. For each mode from I to XI, the indicator $|\Delta f| \times \sqrt{|v_g(k_z > 0)|}$ calculated as a function of $l$ (10 nm ≤ $l$ ≤ 600 nm, where the interval of $l$ is 10 nm) and $k_z$ (0.1 rad/μm ≤ $k_z$ ≤ 10 rad/μm, where the interval of $k_z$ is 0.1 rad/μm) is introduced. The square root is used to moderate the rapid increase of $v_g$ at low wavenumbers and to keep $|\Delta f|$ to be greater than approximately 1 GHz. The calculation results are depicted in Fig. 5, where the size of each circle represents $v_g$ (specific values are also displayed). The color bar expresses $|\Delta f|$ when $|\Delta f| \times \sqrt{|v_g(k_z > 0)|}$ is the maximum value.

The interplay of increasing the bilayer thickness and using a higher-order mode results in sizable nonreciprocity, accompanied by a higher group velocity, whereas the frequency difference decreases. It should be noted that only one spin-wave mode (Mode I, II, III, etc.) is considered in the calculation of $|\Delta f|$ depicted in Fig. 5. Because the frequency gap of the adjacent modes becomes narrow as the thickness increases (Figs. 1(c) and 1(d)), the nonreciprocity observable when only one mode is considered is suppressed because of the presence of other modes, especially in the large thickness region. Therefore, there is a trade-off between achieving a significant frequency difference and maintaining a high group velocity, necessitating the optimal selection of the bilayer thickness for practical applications.

## V. CONCLUSION

Spin-wave nonreciprocity at low wavenumbers in a FM bilayer was theoretically investigated and experimentally demonstrated. In the previous study, nonreciprocity was larger in the high-wavenumber regime and spin waves at low wavenumbers were difficult to excite efficiently using the BLS measurement in the FM bilayer. Thus, nonreciprocity in a low-wavenumber regime has so far received little attention. Furthermore, because the DE mode in the regime of low wavenumbers exhibits a high group velocity, realizing large nonreciprocity at low wavenumbers has been sought for practical applications in spin-wave-based devices.

Consequently, we found that spin waves exhibit large nonreciprocity even in a low-wavenumber region, leading to a high group velocity, by increasing the thickness of the bilayer. This increase is attributed to the decrease in the frequencies of the PSSW modes as the thickness increases, resulting in the coupling between the DE mode and the PSSW modes at low wavenumbers. We also optimized the thickness of the bilayer using the theoretical model, which indicates that spin waves exhibit nonreciprocity with a high group velocity through an increase in thickness and the use of higher-order modes. These findings establish a significant design principle for the development of high-speed and low-loss spin-wave devices.

**APPENDIX: DETAILS OF THE THEORETICAL CALCULATION**

An FM bilayer composed of two FM layers with identical thickness is considered, where the external magnetic field and wavevector are oriented in the DE geometry (Fig. 1(a)). In this configuration, the LL equation can be represented as

$$i\omega \boldsymbol{\eta}(x) = -\gamma \mu_0 H_0 \boldsymbol{\eta}(x) \times \hat{\boldsymbol{y}} + \gamma \frac{2\langle A \rangle}{\langle M_s \rangle}\left(\frac{\partial^2}{\partial x^2} - k_z^2\right)\boldsymbol{\eta}(x) \times \hat{\boldsymbol{y}}$$

$$-\gamma \mu_0 \langle M_s \rangle \hat{\boldsymbol{y}} \times \int_{-\frac{l}{2}}^{\frac{l}{2}} dx'\, G_{k_z}(x-x')\{1+a(x')\}\boldsymbol{\eta}(x')$$

$$+\gamma \mu_0 \hat{\boldsymbol{z}} \times \frac{2}{\mu_0 M_s(x)}\left\{K_s^{\text{bot}}\delta\left(x+\frac{l}{2}\right)\eta_x\left(-\frac{l}{2}\right) + K_s^{\text{top}}\delta\left(x-\frac{l}{2}\right)\eta_x\left(\frac{l}{2}\right)\right\} \quad \text{(A1)}$$

where $\omega$ is the angular velocity, $\boldsymbol{\eta}(x)$ is the normalized amplitude vector as a function of $x$, $\gamma$ is the gyromagnetic ratio, $\mu_0$ is the vacuum permeability, $H_0$ is the external magnetic field, $\hat{\boldsymbol{y}}$ and $\hat{\boldsymbol{z}}$ are the unit vectors in the $y$ and $z$ directions, $G_{k_z}(x)$ is the magnetostatic Green's function [23] and is written

as

$$G_{k_z}(x) = \begin{pmatrix} G_{k_z}^{xx} & G_{k_z}^{xz} \\ G_{k_z}^{zx} & G_{k_z}^{zz} \end{pmatrix} = \begin{pmatrix} \frac{|k_z|}{2} e^{-|k_z x|} - \delta(x) & -\frac{ik_z}{2} \mathrm{sgn}(x) e^{-|k_z x|} \\ -\frac{ik_z}{2} \mathrm{sgn}(x) e^{-|k_z x|} & -\frac{|k_z|}{2} e^{-|k_z x|} \end{pmatrix} \quad (A2)$$

and $a(x)$ can be written as $(M_{s1} - M_{s2})\,\mathrm{sgn}(x)/(M_{s1} + M_{s2})$. The second and third terms on the right-hand side of Eq. (A1) correspond to an exchange field and dipolar field [14], and the fourth term represents the effect of a uniaxial out-of-plane surface anisotropy [11]. When considering components from 0 to $n$, $\boldsymbol{\eta}(x)$ is expressed as

$$\boldsymbol{\eta}(x) = \left( \sum_{q=0}^{n} \eta_x^q S_q(x), \sum_{q=0}^{n} \eta_z^q S_q(x) \right)^{\mathrm{T}} \quad (A3)$$

using Eq. (1). Substituting Eq. (A3) into Eq. (A1) and applying $\int_{-l/2}^{l/2} dx\, S_p(x)$, Eq. (A1) is rewritten as $i\Omega\boldsymbol{\eta} = C\boldsymbol{\eta}$, where $\Omega = \omega/(\gamma\mu_0\langle M_s\rangle)$. The explicit form of $C$ is derived in Supplemental Material B [22].

**Acknowledgments**

This work is supported in part by a Grant-in-Aid for Scientific Research (B) (Grant No.21H01798) and the JSPS Research Fellow Program (Grant No. 22KJ1956). We thank Sotaro Mae and Kento Yasui for their advice on the measurement setup and device fabrication process.

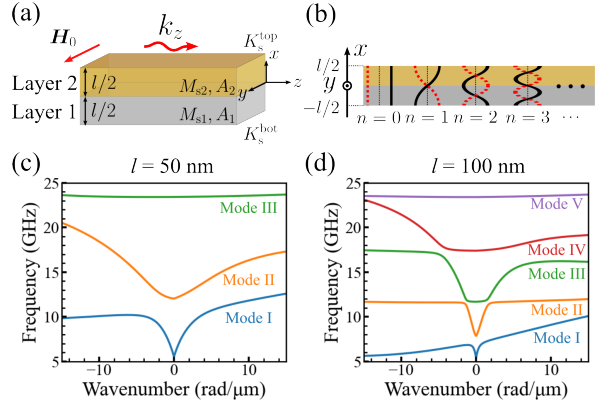

FIG. 1. (a) Schematic of the FM bilayer. The external magnetic field and wavenumber are arranged in the DE geometry. (b) Orthonormal basis functions of the spin waves, where $n$ denotes the number of nodes. (c) and (d) illustrate the dispersion relations for an FM bilayer with thicknesses of 50 nm and 100 nm. The DE mode and PSSW modes exhibit mode repulsion at the crossing points, and the hybridized modes are labeled as Mode I, Mode II, and so on.

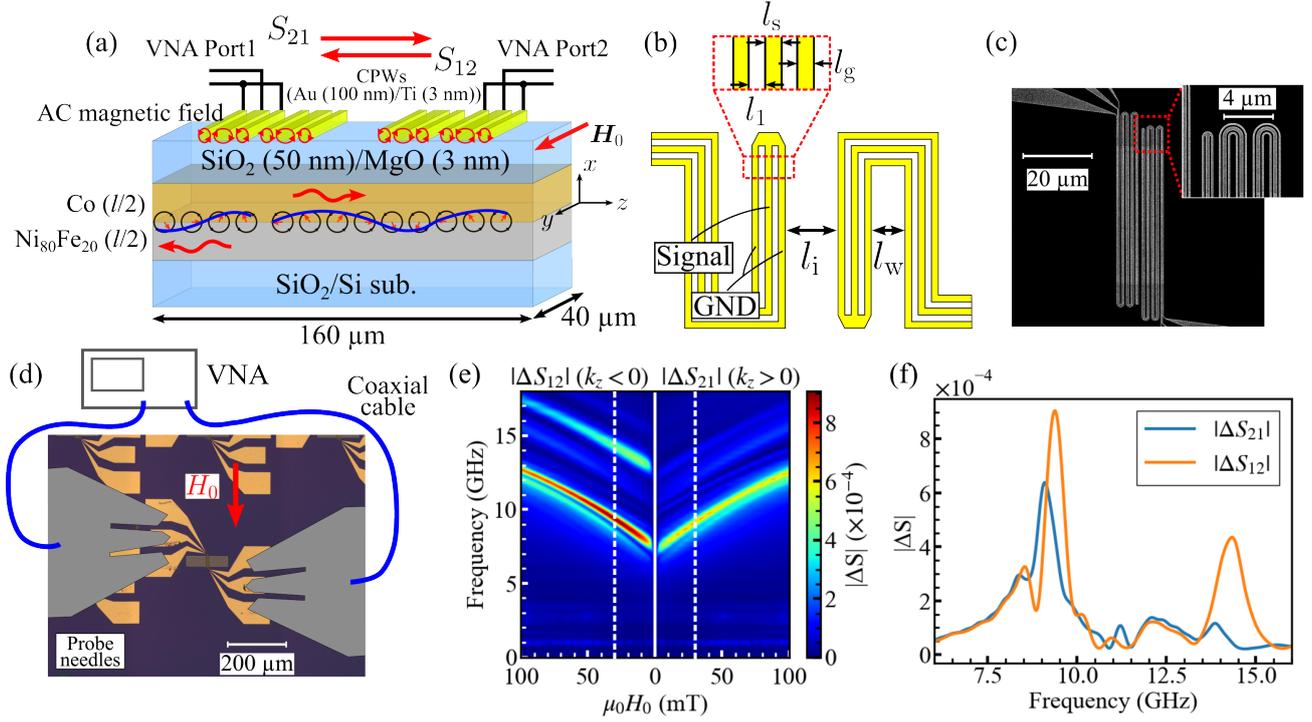

FIG. 2. (a) Schematic of the device structure. A total of twelve devices is fabricated (six types of CPWs listed in Table I are equipped on Co (25 nm)/Ni$_{80}$Fe$_{20}$ (25 nm) or Co (50 nm)/Ni$_{80}$Fe$_{20}$ (50 nm)). (b) Illustration of a meander-shaped CPW. (c) SEM image of CPW6, which has the narrowest signal or ground lines of the six types of CPWs. (d) Schematic representation of the measurement setup. (e) Measurement result of Co (25 nm)/Ni$_{80}$Fe$_{20}$ (25 nm) sample equipped with CPW4 (after noise removal). (f) $|\Delta S_{21}|$ and $|\Delta S_{12}|$ as functions of frequency at $\mu_0 H_0$ = 30 mT.

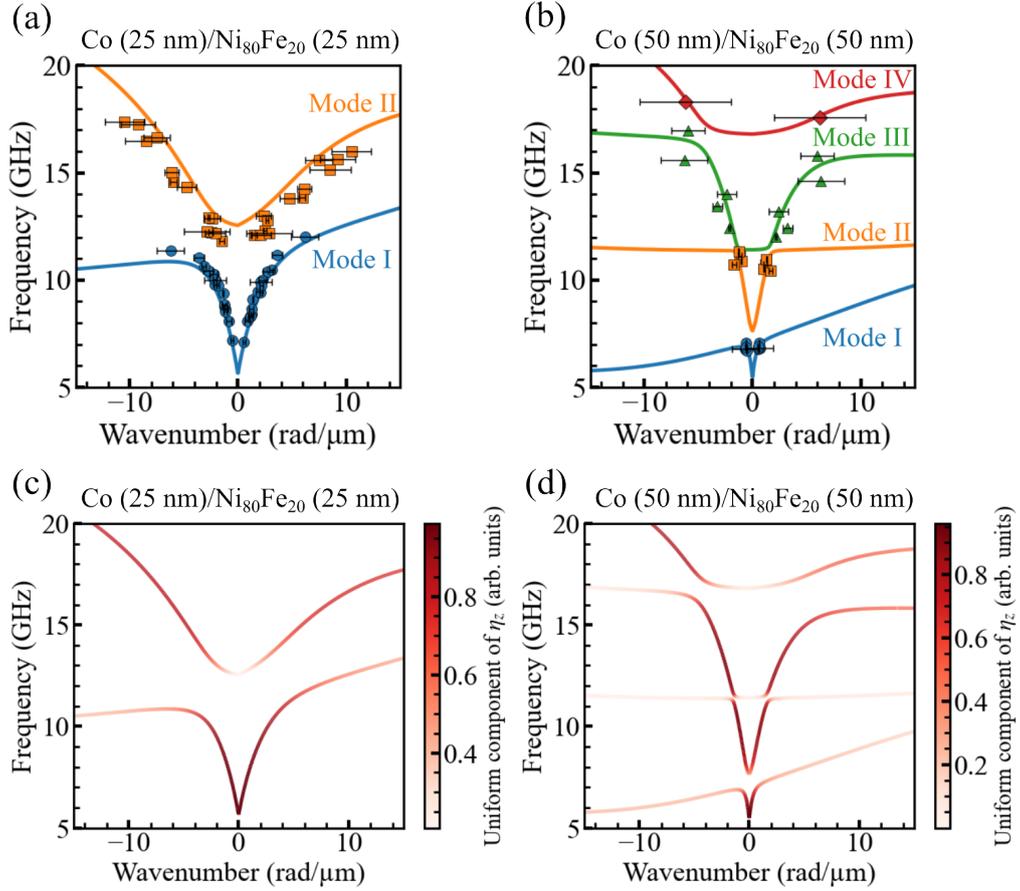

FIG. 3. Resonant frequencies of each wavenumber and mode obtained from the fitting are plotted with the theoretical curves for (a) Co (25 nm)/Ni$_{80}$Fe$_{20}$ (25 nm) (bilayer A) and (b) Co (50 nm)/Ni$_{80}$Fe$_{20}$ (50 nm) (bilayer B). (c) and (d) represent the amplitude of the uniform ($n = 0$) magnetization component in the $z$ direction, $|\eta_z^0|$, calculated using the theoretical model of Eq. (A1).

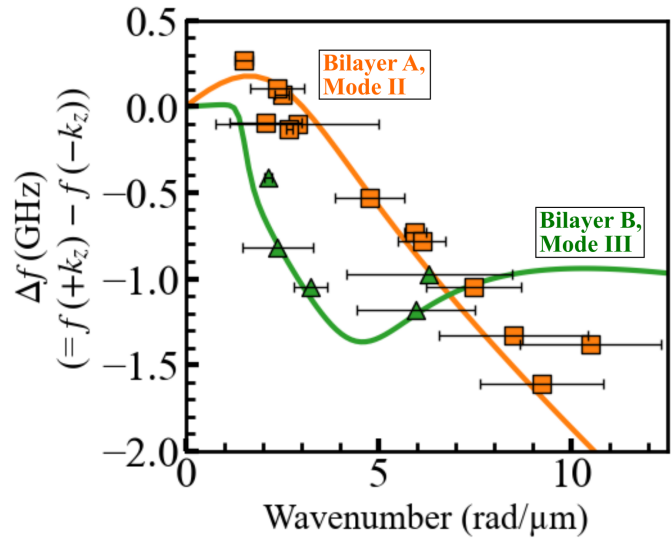

FIG. 4. The frequency differences $\Delta f$ of Mode II in Fig. 3(a) and Mode III in Fig. 3(b) are compared as functions of the wavenumber. The dots represent fitting results, while the curves are calculated using the theoretical model.

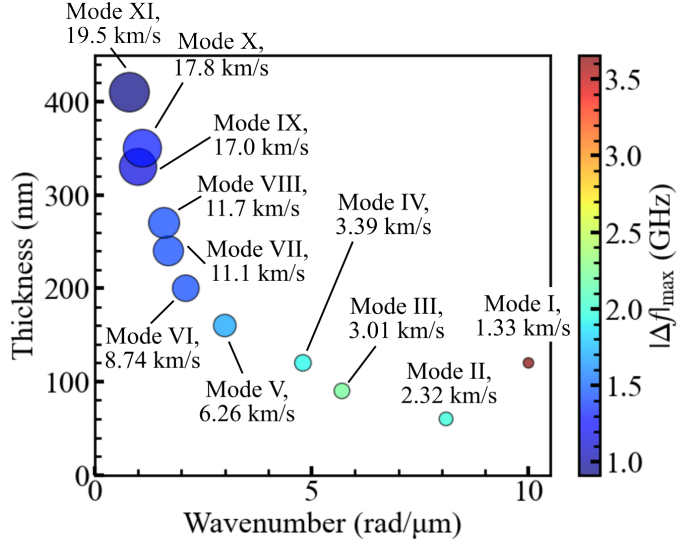

FIG. 5. Calculated results of the maximum $|\Delta f| \times \sqrt{v_g(k_z > 0)}$ as a function of the bilayer thickness and wavenumber for each mode. The size of each circle represents $v_g$, and the color bar indicates the value of $|\Delta f|$ when $|\Delta f| \times \sqrt{v_g(k_z > 0)}$ is the maximum value.

TABLE I. List of the prepared CPWs

|  | $l_s$ ($\mu$m) | $l_1$ ($\mu$m) | $l_g$ ($\mu$m) | $l_w$ ($\mu$m) | $l_i$ ($\mu$m) | Number of meanders |
|---|---|---|---|---|---|---|
| CPW1 | 2 | 1 | 2 | 1 | 1.5 | 1 |
| CPW2 | 1.3 | 0.65 | 0.65 | 1 | 1.5 | 2 |
| CPW3 | 1 | 0.5 | 0.5 | 0.8 | 1.5 | 3 |
| CPW4 | 0.6 | 0.3 | 0.3 | 0.8 | 1.5 | 4 |
| CPW5 | 0.5 | 0.25 | 0.25 | 0.7 | 1.5 | 4 |
| CPW6 | 0.3 | 0.15 | 0.15 | 0.5 | 1 | 4 |